\documentclass[preprint,twocolumn,tightenlines,10pt,prl]{revtex4}%
\usepackage{amsfonts}
\usepackage{amsmath}
\usepackage{amssymb}
\usepackage{graphicx}%
\begin{document}
\title{Observation of Discrete Energy Levels in a Quantum Confined System}
\author{L. L. A. Adams, B. W. Lang and A. M. Goldman}
\affiliation{School of Physics and Astronomy, University of Minnesota, 116 Church St. SE,
Minneapolis, MN 55455, USA}

\pacs{PACS number}

\begin{abstract}
Low temperature scanning tunneling microscope images and spectroscopic data
have been obtained on subnanometer size Pb clusters fabricated using the
technique of buffer layer assisted growth. Discrete energy levels were
resolved in current-voltage characteristics as current peaks rather than
current steps. Distributions of peak voltage spacings and peak current heights
were consistent with Wigner-Dyson and Porter-Thomas distributions
respectively, suggesting the relevance of random matrix theory to the
description of the electronic eigenstates of the clusters. The observation of
peaks rather than steps in the current-voltage characteristics is attributed
to a resonant tunneling process involving the discrete energy levels of the
cluster, the tip, and the states at the interface between the cluster and the
substrate surface.Shell document for REV\TeX{} 4.

\end{abstract}
\volumeyear{year}
\volumenumber{number}
\issuenumber{number}
\received[Received text]{date}

\revised[Revised text]{}

\startpage{1}
\endpage{102}
\maketitle

Random matrix theory (RMT) \cite{RMT}\ is believed to describe the
distribution of discrete energy states of quantum systems whose underlying
classical behaviors are chaotic. In particular the Wigner-Dyson \cite{WD} and
Porter-Thomas distributions \cite{PT} describe the level spacings and
probability densities of the eigenfunctions, respectively. Features of Coulomb
staircase tunneling characteristics have provided evidence of discrete energy
levels in clusters of Au \cite{Wang}, InAs \cite{InAs}, and CdSe
\cite{CdSe}\ studied using scanning probe techniques, and in Al grains
\cite{Ralph} in fixed tunneling geometries.\ Peaks rather than steps have been
found in quantum well geometries involving tunneling between electrodes, each
characterized by discrete energy levels \cite{Chang}-\cite{van der Vaart}. In
this Letter, we report the observation of peaks in the $I-V$ characteristics
of pancake shaped Pb clusters investigated using scanning probe
techniques.\ We interpret these peaks as evidence of discrete energy levels.
We find that the distribution of peak spacings, measured at different
positions on the cluster, and which we associate with the distribution of
energy levels spacings, $\Delta$, are consistent with the predictions of
random matrix theory (RMT) in particular the orthogonal Wigner-Dyson
distribution relevant to systems exhibiting time-reversal invariance symmetry.

\ \ It has also been possible to carry out a statistical analysis of values of
the current found at these various peaks. The magnitude of the tunneling
current is in part determined by the probability of an electron tunneling from
the scanning tunneling microscope (STM) tip to the cluster.\ The histogram of
current peaks is found to be consistent with Porter-Thomas statistics. The
latter are believed to describe the distribution of probability densities of
eigenfunctions characteristized by random matrices. Our hypothesis regarding
the observation of peaks in the $I-V$ characteristics is that they are a
consequence of a two-step tunneling process, from the tip to the cluster and
then to a state with a narrow, well-defined, energy at the interface between
the cluster and the semiconductor\ substrate.

Cluster fabrication was carried out in an ultra-high vacuum chamber using the
technique of buffer layered assisted growth \cite{Weaver}. This growth
chamber, which was equipped with Knudsen cells, was joined to the vacuum
chamber of a commercial low temperature ultrahigh vacuum scanning tunneling
microscope. Samples could be moved between chambers without breaking vacuum.
Prior to the fabrication of clusters, the native oxide of the Si substrate was
removed using a standard acid etch. Titanium/platinum electrodes were first
deposited onto the substrate through a stainless steel mask using a separate
electron beam evaporation system. The n-type (phosphorous doped) Si(111) wafer
employed as a substrate, was miscut by 0.5$%
{{}^\circ}%
$ and had a room temperature bulk resistivity
$>$
1000 $\Omega-cm$, as specified by Virginia Semiconductor. After heating the
substrate to 400$%
{{}^\circ}%
$C in ultrahigh vacuum for two hours, the substrate was cooled in the
vacuum\ chamber using a flow-through cryostat. This was a two-stage process,
first cooling with liquid nitrogen and then with liquid helium. Once the
substrate temperature fell below 60 K, Xe gas was absorbed on its surface to
produce a four monolayer thick film. Subsequently a Pb film with a nominal
thickness of 0.2 \AA , as measured by a calibrated quartz crystal monitor, was
deposited onto the adsorbed Xe layer. Clusters were then formed on the
substrate by desorbing the Xe layer as the sample temperature gradually
approached room temperature. The sample was then transferred to the STM
chamber without breaking vacuum. \ 

Chemically etched W tips were used to perform the measurements. These were
tested \textit{in situ }by\textit{\ }imaging a cleaved graphite surface and
achieving atomic resolution. This was done prior to obtaining topographical
and spectroscopic information about the clusters. All spectra were obtained
during topographical imaging by interrupting the STM feedback and working in
constant height mode. However $I-V$ characteristics were acquired in two
different ways: either by stopping the scan over a particular site on a
cluster and then sweeping the bias voltage, or automatically acquiring data at
every raster point in a topographical scan. There was no appreciable
difference between sets of data obtained in these two different ways. \ 

The inset in Fig.\thinspace1 displays a topographical trace obtained using a
STM illustrating several clusters on top of a Si substrate. It is important to
note that the clusters that exhibit peaks in their $I-V$ characteristics were
effectively pancake shaped, with diameters on the order of 3 nm and heights
between 0.3 \ and 1.5 nm. The main part of Fig.\thinspace1 shows a typical
plot of the $I-V$ characteristics at a particular location on one of these
clusters. The data were obtained at a temperature of 4.2 K, which is low
enough that sharp features are resolvable. The typical width of a peak was
approximately 5 meV. The cluster's "image" size for this particular example
was 3.7 nm in length, 2.6 nm in width and 0.7 nm in height. (The actual
cluster size is likely to be smaller due to convolution effects associated
with the STM tip. \cite{Weaver})\ To interpret this data the hypothesis that
the peaks resulted from resonant tunneling involving the discrete electronic
energy levels of the cluster was adopted. The results of a statistical
analysis of the data assuming that this is the case, will be described first.
The issue as to why this might be true, and why the energy levels appear as
peaks rather than steps in the $I-V$ characteristic will then be discussed. \ 

In order to obtain statistical data, it is important to note that the observed
$I-V$ characteristics vary with position across a cluster as shown in
Fig.\thinspace2 where the tunneling current obtained at a fixed bias voltage
as a function of position is displayed. By measuring the $I-V$ characteristics
at regular positions on a specific cluster, a large data set of peak positions
and peak heights can be obtained for the given cluster. After acquiring data
at approximately 280 locations on a specific cluster histograms of peak
spacings and peak heights were generated using the program ROOT \cite{root}.

\ The histogram of normalized peak spacings, (normalized to the mean spacing
of each individual trace) was then fit by the distribution function%
\begin{equation}
P(s)\,=\,b_{\beta}\,s^{\beta}\,\exp(-a_{\beta}\,s^{c_{\beta}}) \label{Wigner}%
\end{equation}
Here the normalized mean spacing, $s$, is simply $\Delta\,/\,\langle
\Delta\rangle\,$, with $\Delta$ representing the level spacing and
$\langle\Delta\rangle$ the mean spacing. Equation \ref{Wigner} \ can represent
the orthogonal ($\beta\,=\,1$), unitary ($\beta\,=\,2$) and symplectic
($\beta\,=\,4$) ensembles \cite{Wigner-Dyson} that correspond to processes
with different symmetries. The orthogonal case corresponds to time reversal
symmetry being preserved in the absence of a magnetic field and describes the
results presented here.\ In the statistical analysis of this histogram fits to
Wigner-Dyson, Poisson, Gaussian, and Lorentzian distributions were made. From
the values of $\chi^{2}$(not shown) it is clear that the Wigner-Dyson
distribution provides the best fit to the data with $a_{\beta}\,=\pi
/4,b_{\beta}=\pi/2$ and $c_{\beta}=2$. \ In Fig.\thinspace3, the histogram of
peak spacings for this cluster, showing the Wigner-Dyson and Poisson fits is plotted.

A similar statistical analysis of the histogram of current peak heights was
also carried out. The following form \cite{Porter Thomas},%
\begin{equation}
P(I)=a\,\left(  \frac{I}{<I>}\right)  ^{b}\exp\left[  -c\left(  \frac{I}%
{<I>}\right)  \right]  \label{Porter}%
\end{equation}
was fit to the data, where $I$ is the peak current and $<I>$ the mean peak
current.\ In this analysis, parameters specific to the Porter-Thomas and
Poisson distributions, which were deemed relevant, were used. The
Porter-Thomas distribution (with $a=(2\pi)^{-\frac{1}{2}},\,b=-1/2$ and
$c=1/2)$ provided a somewhat better fit to the data than the Poisson
distribution. Figure 4 shows a plot of the histogram along with curves
associated with the best fits of the Porter-Thomas and Poisson distributions.
The results of this analysis support the interpretation that these
measurements are yielding spectroscopic information relating to the energy levels.

We now turn to the issue of why peaks rather than steps are found in the $I-V$
characteristics of the clusters. It is known that there are interface states
and Fermi level pinning at epitaxial Pb/Si(111) interfaces involving n-type Si
\cite{pinning}. One such state pins the Fermi level just above the valence
band minimum. This state has no measurable dispersion.\ Although the present
configuration is not one in which Pb clusters were grown epitaxially on Si,
one might expect to find an interface state of this sort. The data, involving
peaks in the $I-V$ characteristics rather than steps, would be consistent with
a picture in which electrons tunnel into the cluster from the STM\ tip and out
of the cluster into the nearly dispersionless interface state. The surface of
the substrate is replete with clusters so that there is likely to be a
continuous distribution of conductive interface states, resulting in a
conducting path connecting a particular cluster to the Pt electrodes.

As the voltage between the STM tip (at virtual ground) and the Pt electrodes
on the substrate is varied, the narrow band interface state is tuned through
the various energy levels of the cluster. The measurements are made with the
STM\ tip at a particular position so that tunneling into the cluster involves
electrons at the Fermi surface of the tip with the nearest unoccupied
eigenstate of the cluster. The values of current at the peaks will reflect the
distribution of tunneling probabilities, which are proportional to the square
of the matrix elements for tunneling between the tip and the discrete
eigenstates of the cluster. Although these eigenstates are extended throughout
the cluster, their amplitudes can be a function of position, explaining the
position dependence of the $I-V$ characteristics \cite{chaos image}, and
explaining why varying the position of the tip results in a large set of
curves, with position-dependent peak spacings and peak heights. When taken in
aggregate this data will reflect the statistical properties of the eigenstates
of \ the cluster.\ Tunneling out of the cluster would involve the eigenstate
whose energy is matched with that of the interface state. As will be discussed
below, this need not be the same state as that involved in the "tunneling in"
process.\ The spectrum of discrete energy levels of the clusters would be
explored by changing the voltage between the STM tip and the interface state.

An important issue is the role of charging energy in the proposed two-step
tunneling process.\ Estimates assuming either two or three dimensional
geometries exceed the values of the mean level spacing, suggesting that charge
transport involves\textit{ inelastic }co-tunneling in which electrons tunnel
into the cluster to a particular state and tunnel out from a diffferent one,
with no net charge being added to the cluster during the process
\cite{Feigelman}-\cite{Glazman}. Since in this picture the eigenstate
amplitudes are position dependent, the threshold for resonant tunneling at a
particular location, and thus the voltage of the first peak, depends upon the
energy of that eigenstate relative to the Fermi energy of the tip.\ As a
consequence the voltage at which the first peak is found can be position
dependent, and the relevant physical quantities are the spacings between peaks
rather than the voltages at which they were found. Normalized spacings were
used in the analysis as a matter of convenience. Additional theoretical work
is needed to elaborate on this hypothesis which appears to be central to
understanding this data. \ 

\ An estimate of the mean level spacing can be made using the nearly free
electron model. In both the two-dimensional and three-dimensional cases, the
discrete particle-in-the-box energy levels' estimate is determined from the
size of the particle and the continuum density of states at the Fermi energy.
For a Pb cluster of dimensions 3.7 nm in length, 2.6 nm in width and 0.7 nm in
height, an estimate of the mean level spacing is 22 meV (two-dimensional) and
7 meV (three-dimensional). The experimental measured mean level spacing is
$\langle\Delta\rangle=13.4\,\,meV$ for this cluster. It is noted that these
values are an order of magnitude bigger than the thermal broadening energy
\ at 4.2 K $(3.5$\ $k_{B}T\thicksim1.3\,meV)$ and the energy gap of
superconducting Pb which is $\thicksim1\,meV$. For this reason one would not
expect to observe features in the tunneling characteristic associated with superconductivity.

The clusters which exhibit discrete energy levels described here were all
extremely small and very thin and irregular in shape. They were effectively
two dimensional quantum dot configurations. Although the explanation of the
current peaks presented above is not substantiated by detailed surface and
interface measurements, the distributions of the peak spacings and peak
heights are consistent with expectations for a system exhibiting chaotic
dynamics of discrete energy levels, probed by resonant tunneling.\ This
provides support for the proposed mechanism. The level spacings are governed
by the orthogonal Wigner-Dyson distribution, appropriate to a system in which
time-reversal symmetry is not broken, as would be expected in zero magnetic
field. The intensities of the tunneling current are found to satisfy a
Porter-Thomas distribution.

We acknowledge valuable discussions with Alex Kaminev and Leonid Glazman.\ We
thank Beth Masimore for help with the initial analysis of the data. This work
was supported by the US\ Department of Energy under grant DE-FG02-02ER46004.

\-

\bigskip\-

\textbf{Figure 1}\ Tunneling current versus voltage at T = 4.2 K. Tunneling is
from a tungsten STM tip into a Pb cluster. Inset: 30.0 nm x 30.0 nm image of
Pb clusters grown by a buffer layer assisted growth technique. This image was
obtained using a bias voltage of -3.0 V with a tunneling current of 2.0
$\times$ 10$^{-9}$ A.

\bigskip

\textbf{Figure 2} \ Tunneling current as a function of position in the plane
of a cluster at a fixed applied voltage of 17 meV at T = 4.2 K.

\bigskip

\textbf{Figure 3} \ Histogram of peak spacings. The solid curve is the fit for
the Wigner-Dyson distribution. The dotted line represents the fit for the
Poisson distribution. There are 413 peak spacings that comprise this histogram
normalized to the mean voltage spacing.

\bigskip

\textbf{Figure 4} \ Histogram of peak heights. The solid curve is the fit for
the Porter-Thomas distribution. The dotted line represents the fit for the
Poisson distribution. There are 851 peak heights that comprise the histogram
normalized to the mean peak current.

\clearpage

\noindent

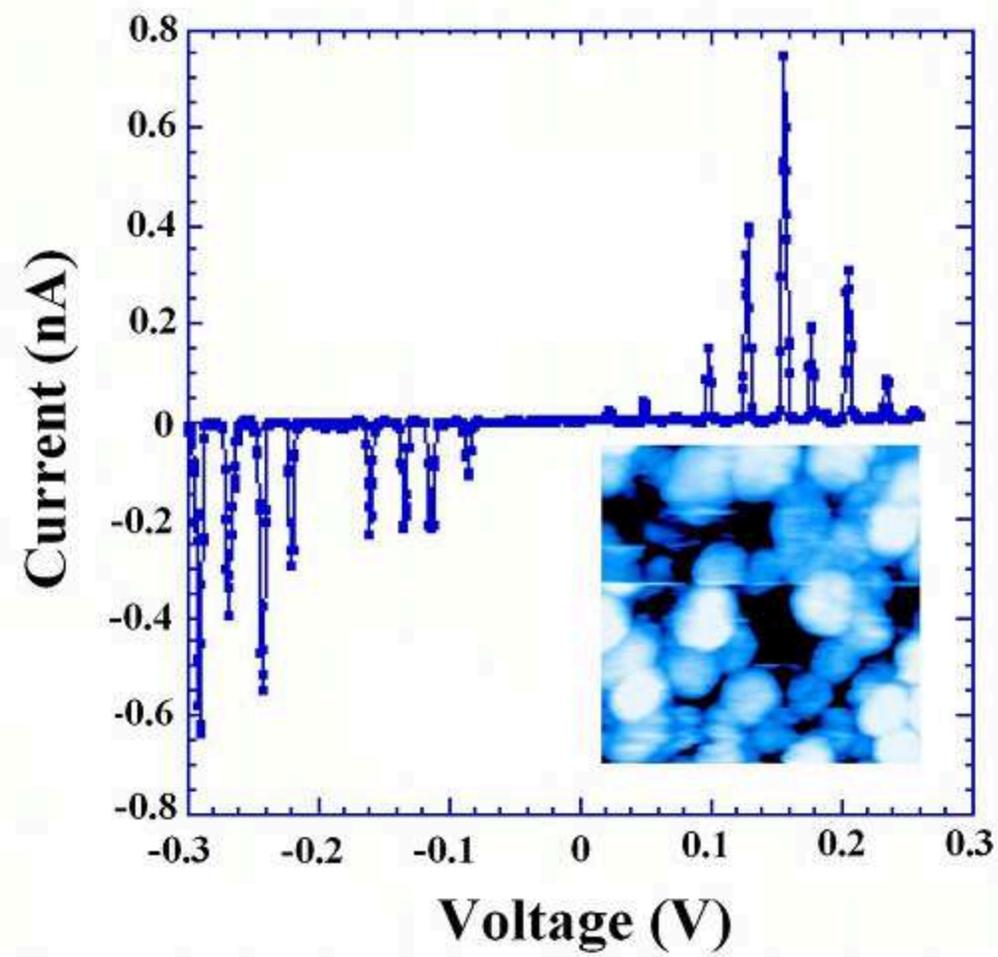
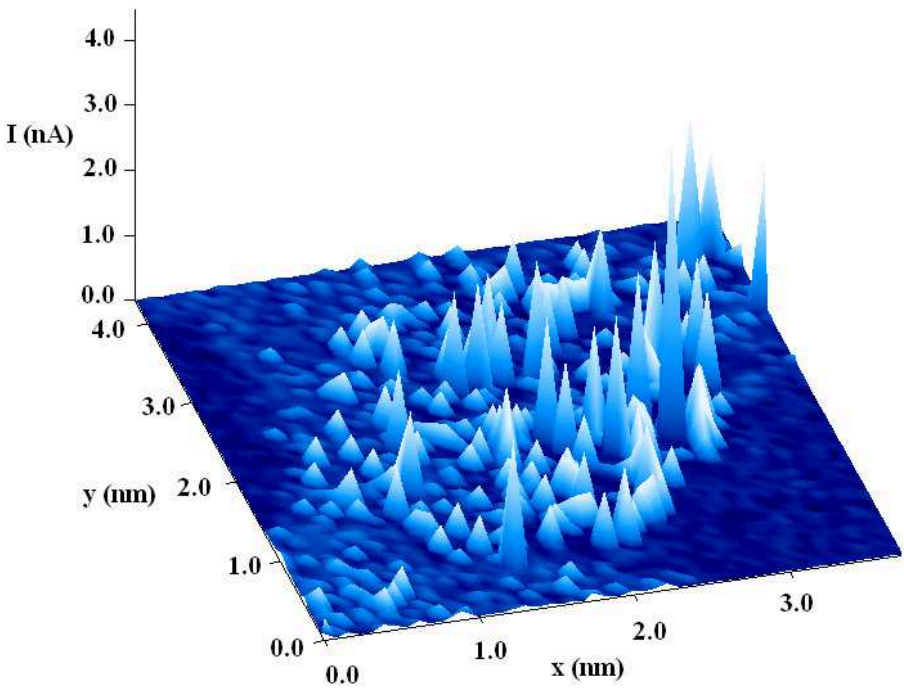
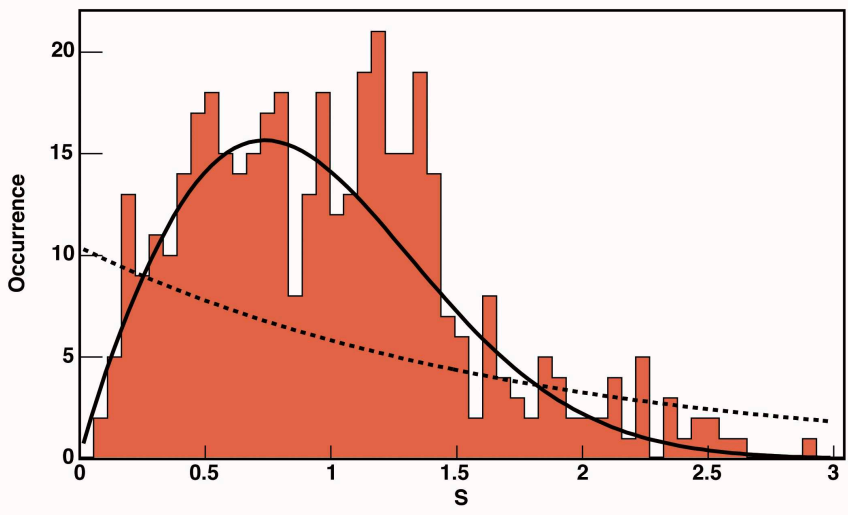
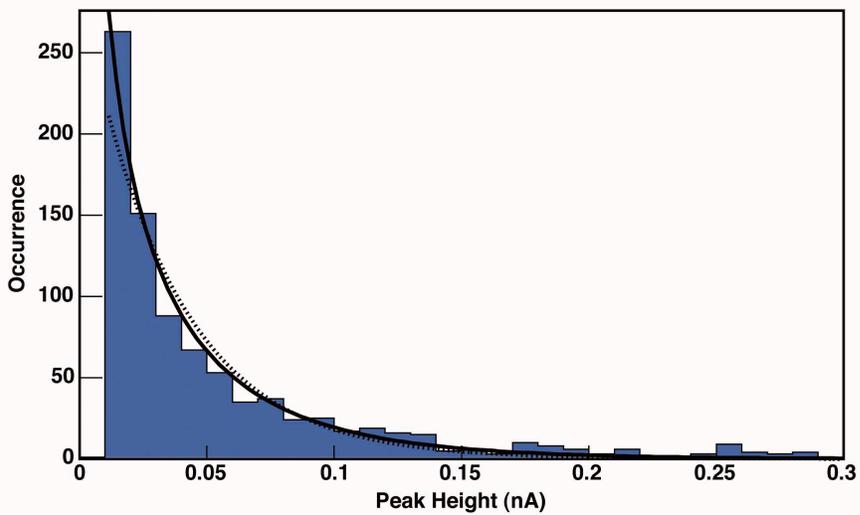
\end{document}